\documentclass[sigconf]{acmart}
\AtBeginDocument{%
  }

\PassOptionsToPackage{table}{xcolor}
\usepackage{colortbl}

\begin{document}

\title{Who Judges the Frame? Auditing Multimodal LLM Judges for News Framing Across Event-Level Perspectives}


\author{Antonela Tommasel}
\orcid{0000-0001-6091-8305}
\affiliation{%
  \institution{Johannes Kepler University Linz}
  \city{Linz}
  \country{Austria}}
\affiliation{%
  \institution{CONICET-UNCPBA, ISISTAN}
  \city{Tandil}
  \country{Argentina}}
\email{antonela.tommasel@jku.at}

\author{Markus Schedl}
\orcid{0000-0003-1706-3406}
\affiliation{%
  \institution{Johannes Kepler University Linz and \\ Linz Institute of Technology}
  \city{Linz}
  \country{Austria}}
\email{markus.schedl@jku.at}

\begin{abstract}
News coverage of major world events is shaped not only by what is reported, but also by how events are framed through text, images and their combination. At the same time, Large Language Models (LLMs), including multimodal LLMs, are increasingly used as scalable instruments for analysing framing, sentiment, ideological slant and perspective differences in multimodal media datasets. This creates a methodological challenge. When used as measurement instruments, LLM outputs may reflect not only content properties, but also model-specific tendencies, prompt design choices, and social, political, cultural, linguistic or modality-specific assumptions.
This work audits LLMs as instruments for large-scale framing and perspective analysis in multimodal news coverage. Using an event-centered dataset of 2025--2026 news coverage, where each event includes left-, center- and right-oriented articles about the same headline, we combine embedding-based measures of within-event viewpoint similarity with model-based assessments of framing constructs across modality-specific and metadata-visible input conditions.  
Rather than treating either dataset labels or model outputs as ground truth, our goal is to examine the usefulness and limitations of LLM-based media analysis. The study contributes an audit protocol that highlights the need to report modality effects, metadata sensitivity, and prompt-induced artifacts alongside substantive claims about news framing and ideological viewpoint differences.

\end{abstract}



\maketitle

\vspace{-0.25cm}\section{Introduction}
Large Language Models (LLMs) are increasingly used as scalable instruments for analysing complex media content. In studies of news, social media, misinformation and political communication, these models can summarize articles, classify frames, infer sentiment, identify actors, compare perspectives, and assess presentation constructs such as emotional intensity, polarization and sensationalism \cite{10.1111/j.1460-2466.2000.tb02843.x, arora-etal-2025-multi, tourni-etal-2021-detecting-frames}. This growing use is motivated by their ability to process large collections of multimodal content without requiring costly human annotation, while applying natural-language definitions to constructs that are difficult to capture with fixed dictionaries or supervised classifiers \cite{arora-etal-2025-multi}. As a result, LLM-based analysis has become an attractive methodological strategy for studying framing and perspective differences in large-scale datasets \cite{arora-etal-2025-multi}.

At the same time, using LLMs as evaluators raises a measurement problem. Their assessments may depend not only on the content being analysed, but also on model-specific tendencies, prompt design choices, and social, political, cultural, linguistic or modality-specific assumptions \cite{wang-etal-2024-large-language-models-fair, chen-etal-2024-humans, VallejoVera2025}. This issue is particularly relevant for multimodal datasets about high-impact public events, where the same event may be represented through different headlines, summaries, images, actors and source contexts. In such settings, LLM-generated judgments should be interpreted as situated outputs of a particular measurement protocol, rather than as neutral observations of framing or ideological perspective.

This work starts from this measurement problem. Rather than assuming that LLMs provide objective judgments of news framing, we examine \textit{how their assessments vary across modality and metadata conditions when applied to event-centered multimodal news coverage}. Our goal is not to replace human perception studies or to establish a definitive ground truth for ideological leaning. Instead, we study the usefulness and limitations of LLM-based media analysis by examining whether the same event-level perspectives produce stable or different model-generated assessments under different input conditions.

We study this problem using a dataset of news events in which each event is associated with left-, center- and right-oriented articles covering the same headline. This event-level structure allows us to compare multiple perspectives on the same event while keeping the underlying topic fixed. We treat the dataset's ideological categories as reference labels for analysis, not as absolute ground truth. 

Our analysis combines two complementary forms of evidence. First, we use textual, visual, and multimodal embeddings to estimate how similar or divergent the three perspectives are within each event and across topics. These representation-level measures provide a non-generative baseline for contextualizing LLM judgments. Second, we use LLM judges to assess selected framing constructs (emotional intensity, conflict framing, polarization framing and sensationalism) through focused Likert-scale prompts. 
We compare these assessments across modality conditions and metadata-visible conditions to examine whether model-generated judgments are shaped by the information made available to the judge. 
Together, these protocols allow us to examine whether LLM-based media analysis produces consistent judgments across different ways of measuring the same event-level perspectives.

Overall, this work offers an audit-oriented study of LLM-based framing and perspective analysis in multimodal news datasets. We argue that LLM-based analysis should not only report what models judge about news content, but also examine how those judgments are shaped by the models and protocols used to produce them.


\section{Experimental Methodology}
This study audits LLM-based framing and perspective analysis in event-centered multimodal news coverage\footnote{The code to reproduce the analysis as well as additional charts can be found at: \url{https://github.com/hcai-mms/judging-frame-multimodal}}. We focus on three research questions:

\textbullet~\textbf{RQ1. Within-event viewpoint divergence.} 
How similar or divergent are left-, center- and right-oriented articles covering the same event in textual, visual and multimodal embedding spaces, and how does this divergence vary across topics?

\textbullet~\textbf{RQ2. Modality sensitivity in framing assessments.} 
How do model-generated assessments of framing constructs change across input modalities? 

\textbullet~\textbf{RQ3. Metadata sensitivity in framing assessments.} 
How do model-generated assessments change when source and ideological reference-label metadata are visible compared with blinded content-only conditions?

Together, these questions separate representation-level differences in the dataset from model-generated judgments about framing. RQ1 provides a non-generative baseline for comparing event-level perspectives, while RQ2 and RQ3 examine whether LLM judgments vary when the same content is presented with different modalities or metadata.

\vspace{-0.3cm}\subsection{Data Collection}
We use the "\textit{Multimodal Understanding Through Impactful World Events}" event-centered dataset of 2025--2026 news coverage\footnote{Instructions to access the dataset can be found at: \url{https://muws-workshop.github.io/cfp/}}. 
Each event is organized around a shared headline and includes topic information together with left-, center- and right-oriented news articles.
We use these ideological categories as reference labels for comparison, not as definitive ground truth about the ideological position of each article.

To support multimodal analysis, we match dataset entries to extracted full articles. Starting from 1,919 original event headlines, full text is available for 2,897 articles associated with 1,651 headlines. 
Among these, 1,268 headlines have matched articles for all three ideological reference categories. 
The final paired multimodal subset contains 915 headlines for which left-, center- and right-oriented articles are all available, and each retained article includes at least one usable image, excluding GIF and WebP files. No preprocessing is applied to either the text or the images.  
For each article in this subset, we use the extracted headline, full body text, source domain, reference ideological category, URL and locally stored images. 
Although some experimental conditions use only textual input, we retain articles with usable images so that modality comparisons are paired across the same set of articles. 

The original dataset associates each article with one of 72 topic labels. These labels include several closely related categories and, in some cases, duplicate labels that differ only by capitalization or minor wording variations (such as \textit{immigration} and \textit{Immigration}). 
To make topic-level analysis more interpretable, we manually merge the original topic labels into ten broader topic groups. The mapping is exhaustive, so every original topic is assigned to exactly one of the ten groups. The full mapping is reported in Table~\ref{tab:topic_mapping}. One group \textit{Other/unclear}, contains three non-substantive labels that appeared only once each in the original distribution. As a result, the articles belonging to this topic were excluded from the analyses.

\begin{table*}[t]
\centering
\footnotesize
\caption{Mapping from original topic labels to merged topic groups.}
\label{tab:topic_mapping}
\vspace{-0.2cm}\begin{tabular}{p{0.28\textwidth}p{0.66\textwidth}}
\toprule
Merged topic group & Original topic labels \\
\midrule

Politics, Elections and Governance &
Politics; Donald Trump; Joe Biden; Elections; Campaign Finance; US Constitution; Supreme Court; Federal State and Tribal Powers; Voting Rights and Voter Fraud; Threats to Democracy; Polarization; Common Ground \\
\addlinespace

Economy, Business, Trade and Housing &
Economy and Jobs; Business; Banking and Finance; Taxes; Trade; free trade agreement; unfair trade practices; Housing and Homelessness \\
\addlinespace

Immigration, Civil Rights and Social Issues &
Immigration; immigration; immigration enforcement; Race and Racism; Civil Rights; Abortion; LGBTQ Issues; Free Speech; Privacy; Sexual Misconduct; Family and Marriage \\
\addlinespace

Crime, Justice and Security &
Criminal Justice; Justice; Defense and Security; Terrorism; Gun Control and Gun Rights; Violence in America \\
\addlinespace

International Relations and Armed Conflict &
Foreign Policy; World; The Americas; China; Middle East; Israel; Russia; Ukraine War; war in Ukraine; Russia-Ukraine war; ongoing conflict with Russia; ongoing war between Russia and Ukraine \\
\addlinespace

Health and COVID-19 &
Healthcare; Public Health; Coronavirus; Life During COVID-19 \\
\addlinespace

Science, Technology, Environment and Weather &
Science; Technology; Energy; Environment; Climate Change; Severe Weather \\
\addlinespace

Media, Information and Public Discourse &
Facts and Fact Checking; Media Industry; Media Bias; Humor and Satire; General News \\
\addlinespace

Culture, Religion, Education and Sports &
Culture; Arts and Entertainment; Religion and Faith; Education; Sports \\
\addlinespace

Other/unclear &
released a report; asking questions; don't approve \\

\bottomrule
\end{tabular}\vspace{-0.3cm}
\end{table*}

\vspace{-0.3cm}\subsection{Model Selection}
For the LLM-based judging experiments, we use \texttt{gemma4:e2b}\footnote{\url{https://ollama.com/library/gemma4:e2b}}, a multimodal LLM accessed through Ollama. The model has 5.12B parameters and was run using the \texttt{Q4\_k\_M} quantized variant. The model fits the main requirements of our study, as it supports multimodal input, allowing the same model to evaluate text-only, image-only, and text-image conditions; it can be run in a controlled local environment; and it is lightweight enough to apply systematically across many events, constructs and input settings. 
Our choice is methodological rather than performance-ma\-xi\-mi\-zing. Larger LLMs may provide stronger general capabilities, but they also introduce practical constraints related to cost, access, version changes and reproducibility. 
As our goal is to audit how a multimodal model behaves as a measurement instrument under different input conditions, a locally runnable LLM provides a controlled setting in which the same prompts, decoding parameters and input variants can be applied consistently. We therefore use the selected model as a controlled case study for auditing modality and metadata sensitivity, rather than as a representative of all possible LLM judges.

All model calls are made with deterministic decoding settings where available, using temperature set to zero. The model receives a fixed system prompt that frames it as a neutral and consistent news-content evaluation assistant and asks it to apply predefined analytical criteria without relying on personal opinions, political assumptions, or external context. Each task requires structured JSON output, which allows for the responses to be parsed and compared across articles, constructs, modalities and metadata conditions.

\vspace{-0.28cm}\subsection{RQ1. Event-level Viewpoint Divergence}

To measure event-level viewpoint divergence, we compute textual, visual, and multimodal embeddings for the left-, center- and right-oriented articles associated with each event. Embeddings are computed using \texttt{nvidia/llama-nemotron-embed-vl-1b-v2}\footnote{\url{https://huggingface.co/nvidia/llama-nemotron-embed-vl-1b-v2}}. Textual embeddings use the article headline and full text, visual embeddings use the article images, and multimodal embeddings combine text and images. When an article contains multiple images, we aggregate them into a single article-level representation. 
For each event and modality, we compute pairwise cosine similarities between the three perspectives: left--center, center--right and left--right. Higher similarity indicates lower viewpoint divergence. We summarize the results using average within-event similarity, left--right similarity, and center asymmetry, defined as the difference between left--center and center--right similarity.

\subsection{RQ2. Modality Sensitivity in Framing Assessments}

To examine modality sensitivity, we evaluate whether model-ge\-ne\-ra\-ted framing scores change depending on the type of input provided to the judge. For each article, we run the same construct-specific evaluation prompt under three input conditions: \textit{text-only} (headline plus full text), \textit{image-only}, and \textit{multimodal} (headline plus full text plus image). 

We assess four framing constructs: emotional intensity, conflict framing, polarization framing and sensationalism. We focus on these four constructs because they capture framing and presentation dimensions that are both theoretically relevant \cite{10.1111/j.1460-2466.2000.tb02843.x} and operationally observable in multimodal news content \cite{arora-etal-2025-multi}. Conflict framing and polarization framing describe whether events are presented through disagreement, opposition, or broader social and ideological division. Emotional intensity and sensationalism capture affective and attention-oriented presentation choices. They are also suitable for multimodal analysis because each can be expressed through language, image selection or the interaction between them.

Table~\ref{tab:prompt_skeleton} summarizes the prompt skeleton used for the individual article-level framing assessments\footnote{The full prompts are available in the accompanying repository.}. 
For each construct, we ask the model to assign a score on a 1--5 Likert scale. Across constructs, scores are interpreted as follows:
1 indicates very low presence of the construct,
2 low presence,
3 moderate presence,
4 high presence, and 
5 very high presence. 
Construct-specific anchors are included in the prompt to distinguish them.

\begin{table*}[t]
\centering
\small
\caption{Prompt skeleton for individual article-level framing assessments. Placeholders are filled according to the construct, input modality, and metadata condition. Source and reference-label fields are included only in metadata-visible conditions.}
\label{tab:prompt_skeleton}
\vspace{-0.2cm}\begin{tabular}{p{0.15\textwidth}p{0.8\textwidth}}
\toprule
Prompt component & Content \\
\midrule

System role &
\textit{You are a neutral and consistent news-content evaluation assistant, designed to apply predefined analytical criteria without relying on personal opinions, political assumptions, or external context. You behave as a controlled measurement instrument for media analysis, prioritizing consistency, evidence from the provided content, and strict adherence to the requested output format.}\\
\addlinespace

Task &
\textit{You are evaluating a news item for one specific framing construct.}

\textit{Your task is not to determine whether the article is biased overall. Your task is only to assess the degree to which the specified construct is present in the provided content.}\\

\addlinespace

Construct specification &
\texttt{\{CONSTRUCT\_NAME\}} and \texttt{\{CONSTRUCT\_DEFINITION\}} are inserted into the prompt. The constructs used in the experiments are emotional intensity, conflict framing, polarization framing, and sensationalism. \\
\addlinespace

Instructions &
\textit{Important instructions:}

\textit{1. Evaluate only the provided content.}

\textit{2. Do not use external knowledge about the event, outlet, author, country, or political context.}

\textit{3. Do not infer or assign the ideological orientation of the article.}

\textit{4. Do not evaluate whether the article is factually true or false.}

\textit{5. Do not evaluate whether the article is good or bad journalism.}

\textit{6. Do not evaluate whether the article is fair, balanced, objective, or biased overall.}

\textit{7. Focus only on observable evidence related to the specified construct.}

\textit{8. If the evidence is ambiguous, limited, or mixed, choose the best-supported score and lower the confidence.}

\textit{9. If there are contradictions or tensions within the provided input, base the score on the full provided input and mention the tension in the evidence field.}

\textit{10. Do not mention information that is not present in the provided content.}\\

\addlinespace

Input condition &
The prompt specifies the input condition using \texttt{\{INPUT\_CONDITION\}}. Depending on the experiment, the model receives article text, images, article text plus images, and optionally source or reference-label metadata. \\
\addlinespace

News item fields &
The prompt includes the available fields for the condition: \texttt{\{ARTICLE\_HEADLINE\}}, \texttt{\{ARTICLE\_TEXT\}}, \texttt{\{IMAGE\_INPUT\}}, and, in metadata-visible conditions, \texttt{\{ARTICLE\_SOURCE\}} and \texttt{\{ARTICLE\_REFERENCE\_LABEL\}}. \\
\addlinespace

Scale &
The model assigns a score on a 1--5 Likert scale, where 1 indicates very low presence of the construct and 5 indicates very high presence. Construct-specific scale anchors are inserted through \texttt{\{SCALE\_ANCHORS\}}. \\
\addlinespace

Output format &
The model is instructed to return only valid JSON with four fields: \texttt{score}, \texttt{confidence}, \texttt{evidence}, and \texttt{uncertainty\_reason}. The evidence field must be grounded only in the provided content. \\

\bottomrule
\end{tabular}\vspace{-0.2cm}
\end{table*}

We compare scores across modality conditions for the same article to measure how much the judgment changes when textual or visual information is added or removed. In particular, we compare text-only with multimodal to estimate the effect of adding images to the article text. We summarize these differences overall, by construct, by reference category, and by merged topic group.

\vspace{-0.35cm}\subsection{RQ3. Metadata Sensitivity in Framing Assessments}
To examine metadata sensitivity, we compare model-generated framing scores when the model receives only content with scores produced when additional metadata is visible. For each article and construct, we run the same evaluation prompt under content-only and metadata-visible conditions. The metadata-visible conditions add source information and headlines’ ideological reference label.

The task, construct definitions, Likert-scale anchors and output format remain the same across conditions. This allows us to isolate whether the inclusion of source or reference-label metadata changes the model’s assessment of the same article. 
We evaluate metadata sensitivity in two ways. First, we compute Spearman correlations between corresponding metadata conditions to assess whether relative score patterns across the same articles are preserved after metadata is added. Second, we compute mean signed score shifts relative to the content-only condition to examine the direction of metadata effects. 
For each article and construct, we compute two signed differences: source-visible minus content-only, and source+label minus content-only. Positive values indicate that the metadata-visible condition increases the model-generated score, while negative values indicate that it decreases the score. We summarize these correlations and signed shifts by construct, modality, and ideological reference category.

\vspace{-0.3cm}\section{Experimental Results}

\subsection{RQ1. Event-level Viewpoint Divergence}

\begin{figure}
    \centering
    \includegraphics[width=0.98\linewidth]{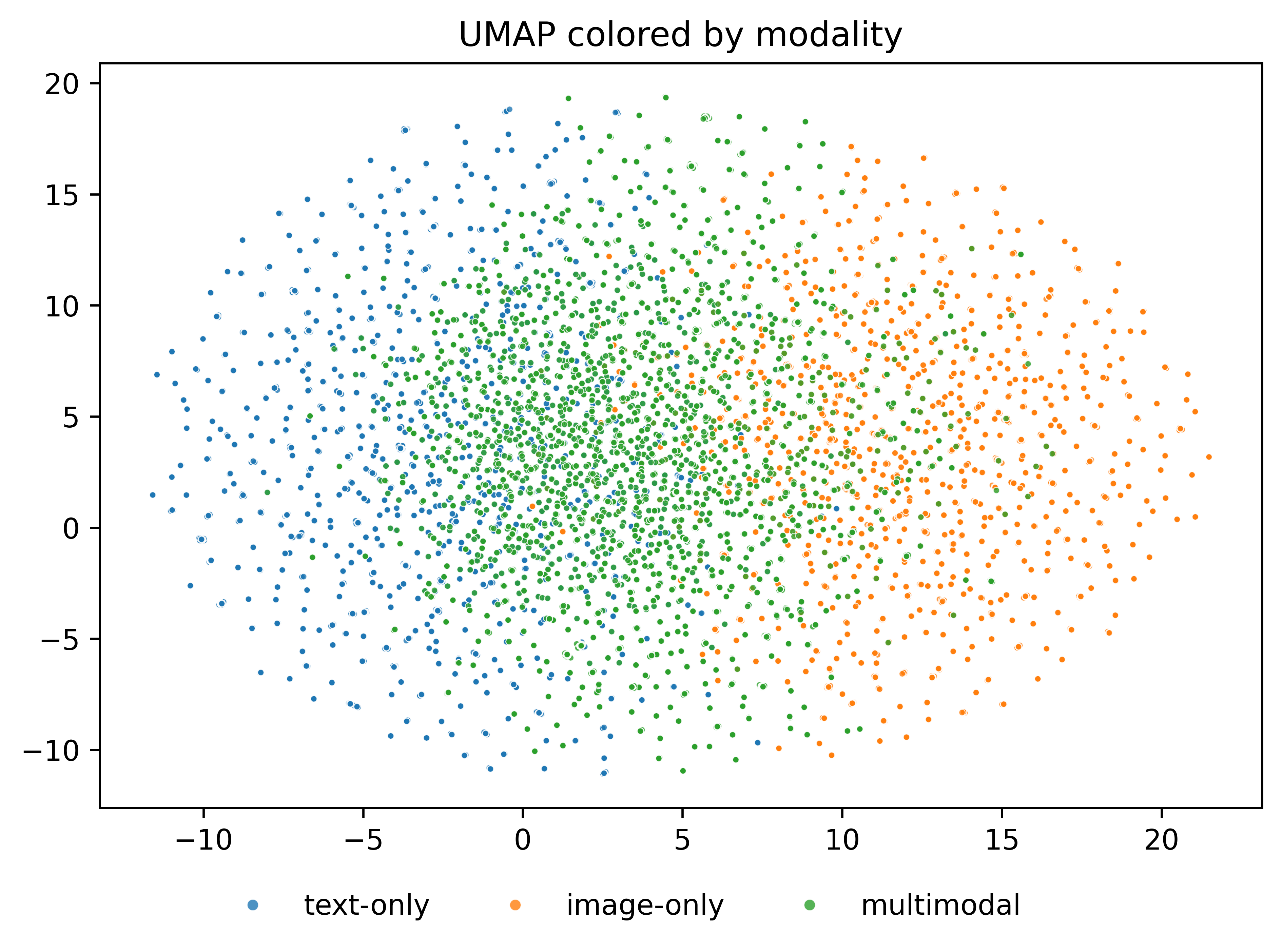}
    \includegraphics[width=0.98\linewidth]{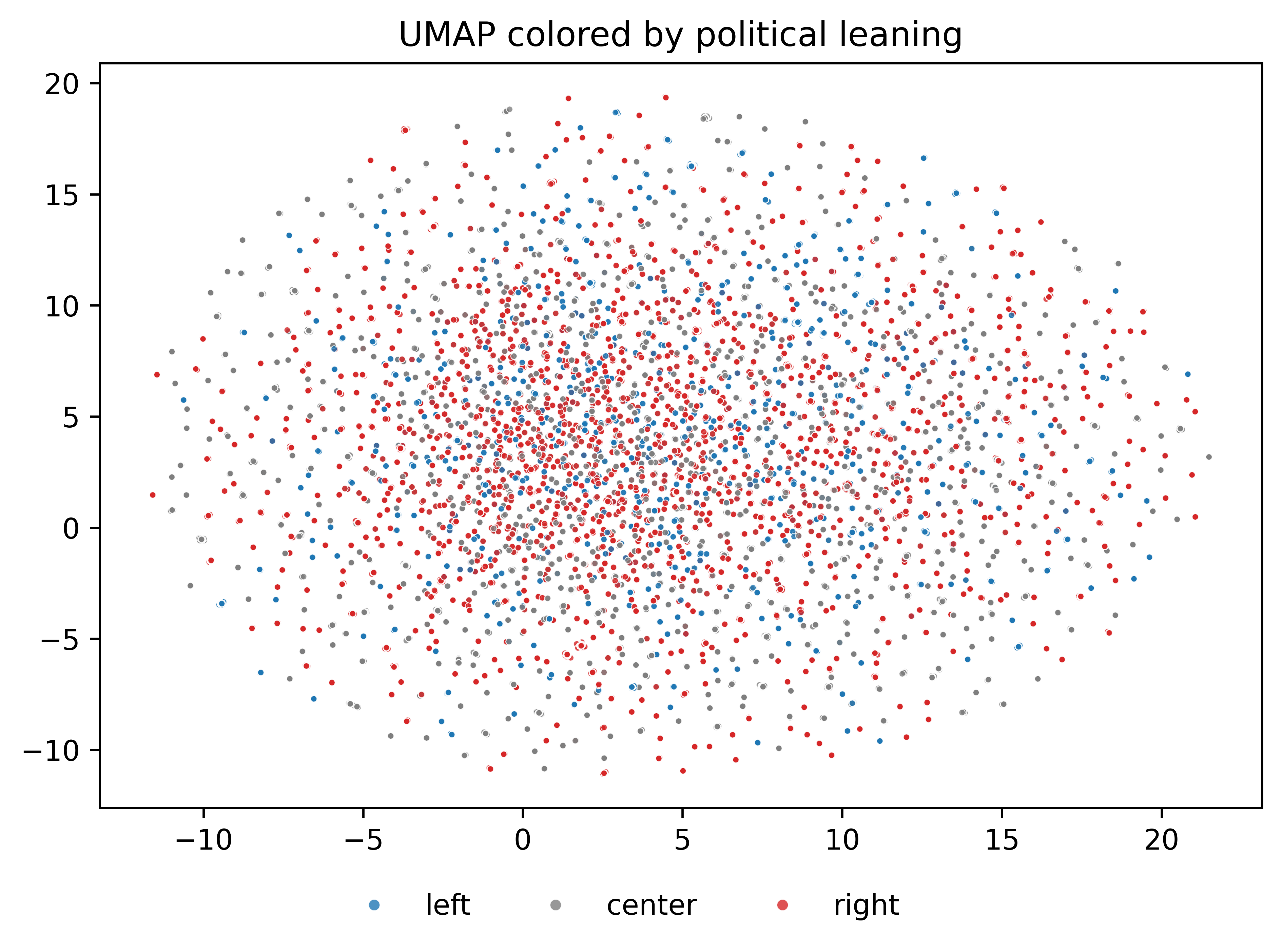}
    \caption{Global structure of article embeddings. UMAP projections of article-level embeddings coloured by input modality and ideological reference category. The projection shows clearer organization by modality than by left-, center- or right-oriented reference category.}
    \label{fig:umap_embeddings}\vspace{-0.25cm}
\end{figure}

We first examine the global structure of the article-level embedding space. Figure~\ref{fig:umap_embeddings} shows UMAP projections \cite{McInnes2018} of the same embeddings coloured by input modality and by ideological reference category. The clearest organization appears by modality, text-based and image-based representations occupy different regions of the projection, while multimodal representations lie between them. In contrast, when the projection is coloured by ideological reference category, left- center- and right-oriented articles are substantially intermixed. This suggests that the reference categories do not form simple global clusters in the embedding space. 

\begin{figure}
    \centering
    \includegraphics[width=1.0\linewidth]{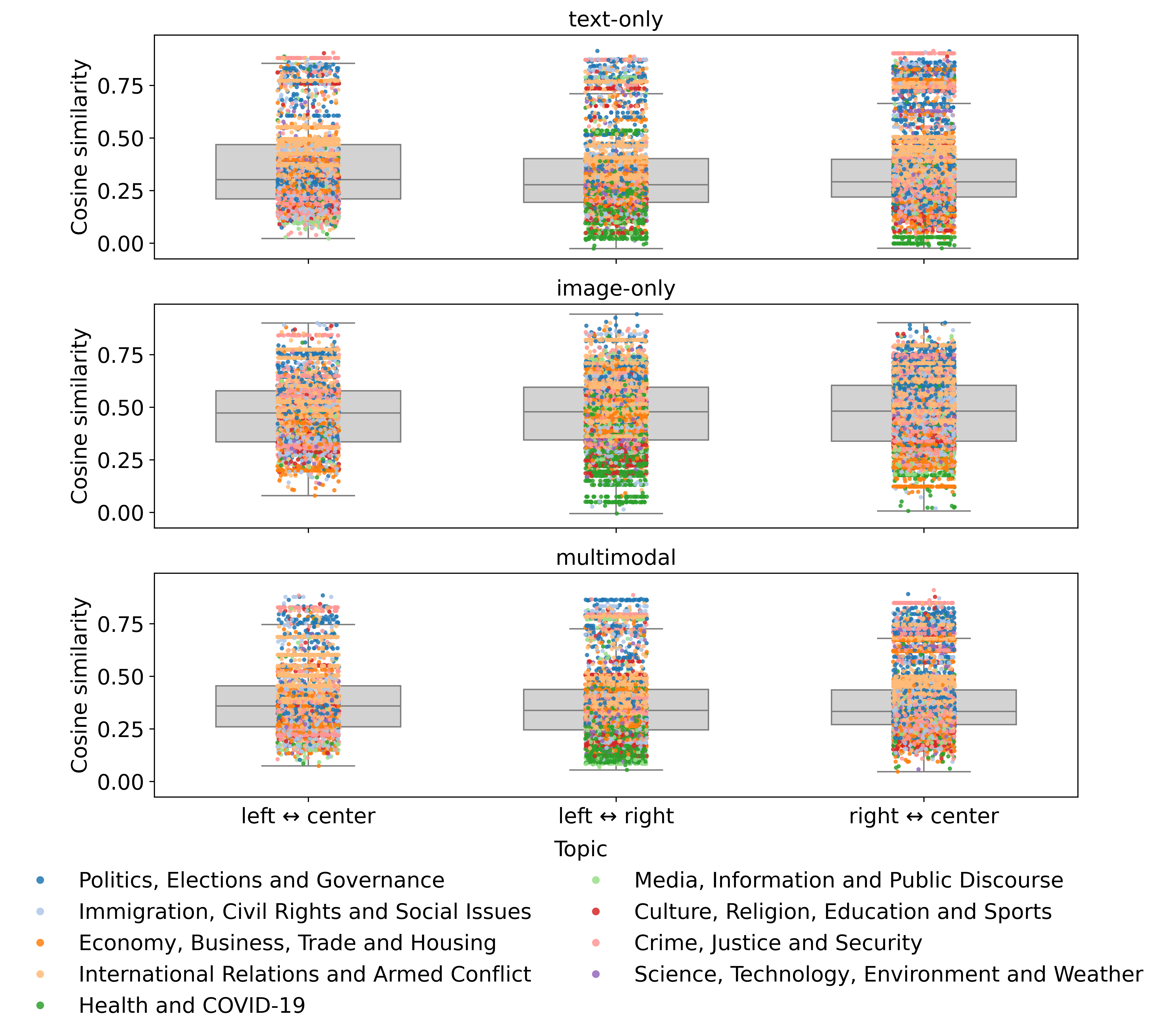}
    \caption{Within-event pairwise similarity modality organized by political leaning. Pairwise cosine similarities are computed between left--center, right--center and left--right articles for the same event. Higher values indicate lower viewpoint divergence. Similarities vary across modality and topic, while the ideological pairs show overlapping distributions.}
    \label{fig:pairwise-similarity}
\end{figure}

\begin{figure*}
    \centering
    \includegraphics[width=0.98\linewidth]{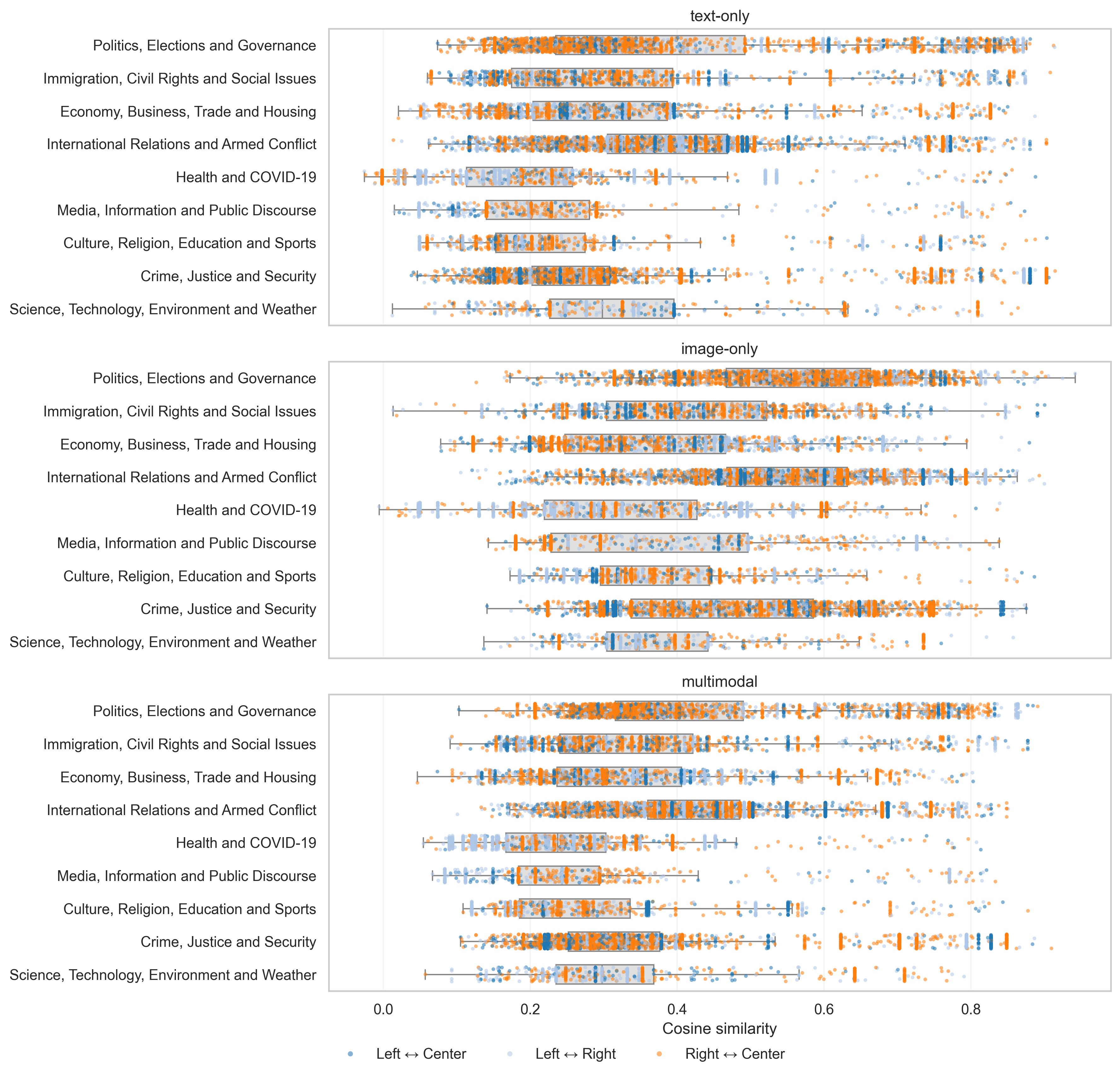}
    \caption{Within-event pairwise similarity by modality, organized by topics. Pairwise cosine similarities are computed between left--center, right--center and left--right articles for the same event. Higher values indicate lower viewpoint divergence.}
    \label{fig:pairwise-similarity2}
\end{figure*}

We then compute the pairwise cosine similarities between the left--center, right--center and left-right articles associated with each event. Higher cosine similarities indicate lower viewpoint divergence. Figure~\ref{fig:pairwise-similarity} and Figure~\ref{fig:pairwise-similarity2} report the within-event similarities across merged topic groups, political leaning and modality conditions. Across topics and modalities, the three ideological pair types show broadly overlapping distributions. In particular, left--right pairs are not consistently less similar than left--center or right--center pairs. This indicates that viewpoint divergence is not organized as a simple linear separation between ideological endpoints.

\vspace{0.15cm}\noindent
Overall, RQ1 shows that articles covering the same event are \textit{often close in embedding space, but their similarity varies substantially across modality and topic}. The strongest global structure is modality-based rather than ideology-based. Within events, \textit{viewpoint divergence is therefore better understood as an event-, topic- and modality-dependent phenomenon}, rather than as a simple global separation between political reference categories.

\subsection{RQ2. Modality Sensitivity in Framing Assessments}

\begin{figure}
    \centering
    \includegraphics[width=0.98\linewidth]{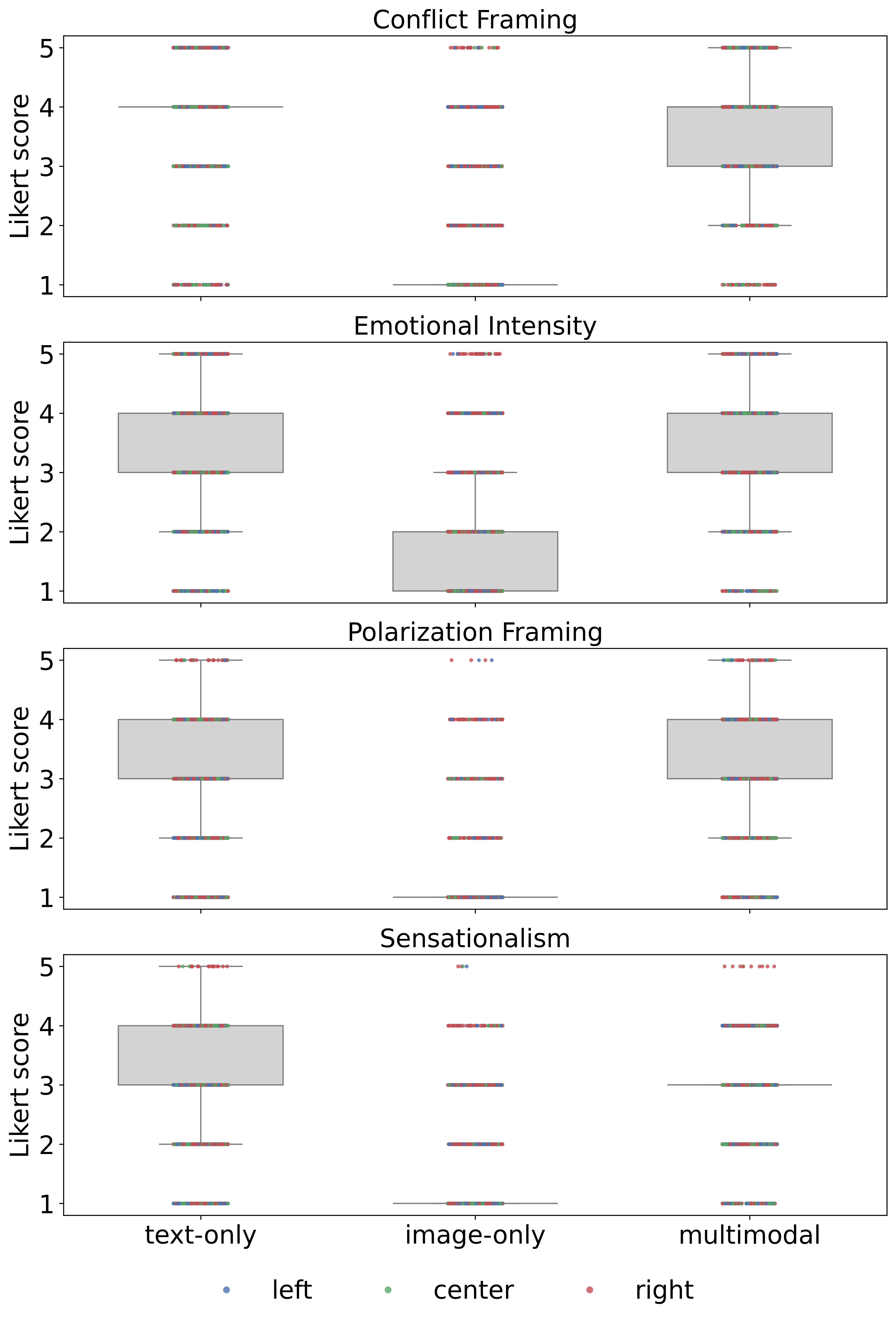}
    \caption{Distribution of model-generated Likert scores for emotional intensity, conflict framing, polarization framing and sensationalism under text-only, image-only and multimodal conditions.}
    \label{fig:likert-distribution}
\end{figure}

We next examine whether model-generated framing assessments change across input modalities. 
Figure~\ref{fig:likert-distribution} shows the distribution of Likert scores by construct and modality. Across all four constructs, text-only and multimodal conditions produce broadly similar score distributions, indicating that adding images to full article text produces limited changes in aggregate scores. In contrast, the image-only condition produces substantially lower scores for most constructs. Conflict framing, emotional intensity and polarization framing tend to receive higher scores when article text is available, while sensationalism receives more moderate scores overall. This pattern suggests that, for the analysed constructs, the multimodal condition is largely anchored in textual content. Images may affect individual cases, especially when visual material provides emotionally salient or dramatic cues, but they do not substantially shift the overall distribution of scores when full text is already available.

We quantify this pattern using both rank agreement and absolute agreement. Table~\ref{tab:rq2_modality_agreement} reports Spearman correlations and intraclass correlation coefficients (ICC) between modality conditions. Spearman correlations between text-only and multimodal are high across constructs, ranging from 0.68 for sensationalism to 0.80 for polarization framing. ICC values show the same pattern, ranging from 0.70 for sensationalism to 0.83 for conflict and polarization framing. This indicates that text-only and multimodal conditions preserve both the relative ordering of articles and their absolute score levels.

In contrast, agreement involving the image-only condition is low. Spearman correlations between text-only and image-only range from 0.11 to 0.19, while ICC values range from 0.02 to 0.04 for conflict framing, polarization framing, and sensationalism, and 0.04 for emotional intensity. Agreement between image-only and headline-text-image is similarly low, with Spearman correlations between 0.12 and 0.17 and ICC values between 0.03 and 0.05. Thus, image-only judgments are not simply lower versions of text-based judgments. The articles scores highly from text are not necessarily the same articles scored highly from images and the absolute Likert score levels also differ.

\begin{table}[t]
\centering
\caption{Agreement between modality conditions for model-generated framing scores. Spearman measures agreement in relative score patterns, while ICC measures agreement in absolute Likert score levels.}
\label{tab:rq2_modality_agreement}
\scalebox{0.83}{\begin{tabular}{lcccccc}
\toprule
 & \multicolumn{2}{c}{Text--Image} & \multicolumn{2}{c}{Text--Multimodal} & \multicolumn{2}{c}{Image--Multimodal} \\
\cmidrule(lr){2-3} \cmidrule(lr){4-5} \cmidrule(lr){6-7}
Construct & $\rho$ & ICC & $\rho$ & ICC & $\rho$ & ICC \\
\midrule
Conflict framing & 0.19 & 0.04 & 0.79 & 0.83 & 0.17 & 0.03 \\
Emotional intensity & 0.15 & 0.04 & 0.74 & 0.77 & 0.15 & 0.05 \\
Polarization framing & 0.11 & 0.02 & 0.80 & 0.83 & 0.12 & 0.03 \\
Sensationalism & 0.12 & 0.04 & 0.68 & 0.70 & 0.15 & 0.05 \\
\bottomrule
\end{tabular}}
\end{table}

\vspace{0.15cm}\noindent
Overall, RQ2 shows clear \textit{modality sensitivity in model-generated framing assessments}. The strongest difference is between image-only and text-containing conditions (i.e., text-only and multimodal). Text-containing  assessments are relatively stable, suggesting that adding images to full text does not substantially change the model’s judgments for the constructs analysed here. However, image-only assessments behave differently, both in absolute Likert score levels and in relative score patterns across articles. These results indicate that \textit{modality conditions should be reported explicitly when using multimodal LLMs for framing analysis, since different input modalities can produce different measurement outcomes}.

\subsection{RQ3. Metadata Sensitivity in Framing Assessments}

We next examine whether model-generated framing assessments change when metadata is made visible to the model. We compare three metadata conditions within the same modality: a content-only condition with no metadata, a source-visible condition, and a source+label condition in which both the source and the headlines' ideological reference label are shown. We focus on same-modality comparisons so that differences can be attributed to metadata visibility rather than to changes in the underlying input modality. 

Descriptive inspection of the score distributions suggests that metadata does not substantially change the aggregate distributions when article text is available. Text-only and multimodal conditions remain broadly similar after adding source or source+label information. The largest visible differences appear between text-containing and image-only conditions, rather than between content-only and metadata-visible variants within the same modality. Image-only assessments are more compressed toward the lower end of the scale, especially for conflict framing and polarization framing, suggesting that the model has limited construct-specific evidence when article text is absent.

\begin{table}[t]
\centering
\caption{Spearman correlations between metadata conditions, computed within the same modality. Higher values indicate greater agreement in relative score patterns after metadata is added.}
\label{tab:rq3_metadata_correlations}
\resizebox{\linewidth}{!}{
\begin{tabular}{llccc}
\toprule
Construct & Modality & No-meta--Source & No-meta--Source+Label & Source--Source+Label \\
\midrule
Conflict framing & text-only & 0.80 & 0.79 & 0.80 \\
Conflict framing & multimodal & 0.82 & 0.81 & 0.77 \\
Conflict framing & image-only & -0.01 & 0.01 & 0.08 \\
\midrule
Emotional intensity & text-only & 0.77 & 0.77 & 0.77 \\
Emotional intensity & multimodal & 0.83 & 0.82 & 0.77 \\
Emotional intensity & image-only & -0.01 & 0.00 & 0.20 \\
\midrule
Polarization framing & text-only & 0.80 & 0.79 & 0.78 \\
Polarization framing & multimodal & 0.84 & 0.84 & 0.78 \\
Polarization framing & image-only & 0.02 & 0.01 & 0.22 \\
\midrule
Sensationalism & text-only & 0.70 & 0.70 & 0.72 \\
Sensationalism & multimodal & 0.77 & 0.76 & 0.74 \\
Sensationalism & image-only & 0.01 & 0.05 & 0.22 \\
\bottomrule
\end{tabular}
}
\end{table}

Table~\ref{tab:rq3_metadata_correlations} reports Spearman correlations between metadata conditions. For text-containing modalities, correlations between content-only and source-visible conditions are high across constructs. In the text-only condition, correlations range from 0.70 for sensationalism to 0.80 for conflict framing and polarization framing. In the multimodal condition, correlations range from 0.77 for sensationalism to 0.84 for polarization framing. Similar patterns hold when comparing content-only with source+label conditions, with correlations ranging from 0.70 to 0.79 for text-only and from 0.76 to 0.84 for multimodal. This indicates that, when article text is available, adding source or source+label metadata generally preserves the relative ordering of articles.

This stability does not hold for image-only inputs. Correlations between content-only and source-visible image conditions are near zero, ranging from -0.01 to 0.02 across constructs. Correlations between content-only and source+label image conditions are also near zero, ranging from 0.00 to 0.05. This suggests that, when the model receives only images, adding metadata substantially changes which articles receive comparatively higher or lower scores. In this setting, source and reference-label information appear to become more influential because the visual content alone provides weaker or less explicit evidence for the framing constructs.

The comparison between source-visible and source+label conditions provides a more focused estimate of the additional effect of showing the reference label after the source is already visible. For text-containing modalities, these correlations remain high, ranging from 0.72 to 0.80 for text-only and from 0.74 to 0.78 for multimodal. For image-only inputs, correlations are higher than in the content-only comparisons but still low, ranging from 0.08 for conflict framing to 0.22 for polarization framing and sensationalism. This suggests that reference labels have limited additional effect when article text is present, but a more noticeable effect when the model has only visual content and source context.

Disaggregating the correlations by ideological reference category does not change the overall pattern. For text-containing modalities, correlations remain high for left-, center- and right-oriented articles, indicating that metadata-visible conditions generally preserve relative score patterns across all three reference categories. However, correlations are slightly lower for center-oriented articles than for left- or right-oriented articles in several text-containing comparisons. For example, averaged across the four constructs and the two text-containing modalities, correlations for content-only versus source-visible conditions are approximately 0.75 for center-oriented articles and 0.81 for both left- and right-oriented articles. In image-only conditions, correlations remain near zero across reference categories when comparing content-only with metadata-visible prompts. This suggests that the main source of metadata sensitivity is the absence of text, although the magnitude of stability can vary somewhat across reference categories.

\begin{table}[t]
\centering
\scriptsize
\setlength{\tabcolsep}{3.5pt}
\renewcommand{\arraystretch}{1.1}
\caption{Mean signed score shifts relative to the content-only condition. 
$\Delta_S$ denotes source-visible minus content-only, and $\Delta_{SL}$ denotes source+label minus content-only. 
Positive values indicate higher model-generated scores after metadata is added. Text-only corresponds to headline plus full text; multimodal corresponds to headline plus full text plus image.}
\label{tab:rq3_signed_shifts}
\vspace{-0.2cm}\begin{tabular}{llrrrrrr}
\toprule
 &  & \multicolumn{2}{c}{Left} & \multicolumn{2}{c}{Center} & \multicolumn{2}{c}{Right} \\
\cmidrule(lr){3-4}\cmidrule(lr){5-6}\cmidrule(lr){7-8}
Construct & Modality 
& $\Delta_S$ & $\Delta_{SL}$ 
& $\Delta_S$ & $\Delta_{SL}$ 
& $\Delta_S$ & $\Delta_{SL}$ \\
\midrule

Conflict framing & Text-only 
& -0.00 & +0.02 & +0.01 & -0.05 & +0.01 & -0.02 \\
Conflict framing & Multimodal 
& +0.09 & -0.06 & +0.04 & -0.10 & +0.04 & -0.04 \\
Conflict framing & Image-only 
& -0.38 & -0.38 & -0.12 & -0.14 & -0.47 & -0.48 \\

\midrule
Emotional intensity & Text-only 
& -0.01 & -0.04 & -0.00 & -0.05 & -0.03 & -0.03 \\
Emotional intensity & Multimodal 
& +0.02 & -0.00 & +0.04 & -0.04 & +0.03 & +0.01 \\
Emotional intensity & Image-only 
& -0.69 & -0.71 & -0.54 & -0.54 & -0.87 & -0.90 \\

\midrule
Polarization framing & Text-only 
& -0.04 & +0.00 & -0.01 & -0.08 & -0.05 & +0.02 \\
Polarization framing & Multimodal 
& +0.03 & -0.03 & +0.02 & -0.06 & +0.04 & +0.01 \\
Polarization framing & Image-only 
& -0.15 & -0.16 & -0.08 & -0.08 & -0.27 & -0.27 \\

\midrule
Sensationalism & Text-only 
& -0.05 & -0.05 & -0.08 & -0.16 & -0.01 & -0.00 \\
Sensationalism & Multimodal 
& -0.07 & -0.05 & -0.05 & -0.04 & -0.06 & -0.02 \\
Sensationalism & Image-only 
& -0.28 & -0.32 & -0.15 & -0.16 & -0.42 & -0.41 \\

\bottomrule
\end{tabular}\vspace{-0.3cm}
\end{table}

Table~\ref{tab:rq3_signed_shifts} reports mean signed score shifts relative to the content-only condition. For text-containing modalities, signed shifts are generally small across constructs and reference categories. In particular, adding source+label metadata does not consistently increase conflict or polarization scores for right-oriented articles. The shifts are -0.02 and -0.04 for conflict framing in the text-only and multimodal conditions, and +0.02 and +0.01 for polarization framing.
A similar pattern holds for left-oriented articles. Adding source+label metadata produces only small shifts in text-containing conditions. The shifts are +0.02 for conflict framing and 0.00 for polarization framing in the text-only condition, and -0.06 and -0.03 in the multimodal condition. Emotional intensity and sensationalism also show small negative or near-zero shifts when text is available.
Conversely, in image-only conditions, metadata-visible prompts reduce scores more substantially for left-oriented articles as well, with source+label shifts of -0.38 for conflict framing, -0.71 for emotional intensity, -0.16 for polarization framing, and -0.32 for sensationalism. 
Thus, the signed-shift analysis does not show a simple ideological-label effect in which left- or right-oriented labels systematically raise particular framing scores. Instead, the strongest directional changes occur in image-only conditions, where metadata generally lowers scores relative to the content-only baseline, with the largest decreases often appearing for right-oriented articles and smaller decreases for center-oriented articles.

\vspace{0.15cm}\noindent
Overall, RQ3 shows that \textit{metadata sensitivity depends strongly on modality}. For text-containing modalities, model-generated framing assessments are relatively stable after adding source or ideological reference-label metadata. For image-only, the same metadata additions substantially alter the relative score patterns assigned to the same articles. Thus, metadata does not uniformly dominate the model’s judgments, but it becomes more consequential when the content provides limited construct-specific evidence. These results indicate that \textit{source and re\-fe\-ren\-ce-label visibility should be treated as part of the measurement protocol and reported explicitly in LLM-based framing analysis}.

\section{Conclusions}
This paper shows that LLM-based framing analysis is sensitive to how news content is represented and what contextual information is made available to the model. 
Across the embedding and judgment experiments, modality emerged as the strongest source of variation: article representations were organized more clearly by input modality than by ideological reference category, and model-generated framing scores were more stable between text-only and multimodal inputs than between text-containing and image-only inputs. Metadata sensitivity was more limited when text was available, but became more consequential in image-only conditions, where source and reference-label information appeared to play a larger role in shaping relative score patterns across article modalities.
Overall, these results show that LLM-generated framing assessments are input-condition dependent, so modality and metadata visibility should be reported as part of the measurement setup.

These findings should be interpreted as a \textit{focused audit of one multimodal LLM judge applied to one event-centered dataset}. 
The study deliberately uses a controlled setting in which the same model, prompts and input conditions are applied systematically. 
This design supports a detailed analysis of modality and metadata sensitivity, but it also limits the scope of the conclusions. 
In particular, the image-only results should be interpreted as the behaviour of the used LLM under our prompt and execution settings, not as a general claim about all multimodal LLM judges. 
Similarly, the representation-level analysis uses a sigle multimodal embedding model, so the embedding-space findings should be understood as model-dependent rather than as properties of the dataset alone.

The study also does not test prompt robustness across alternative prompt formulations, and the model-generated Likert scores are not validated against human or expert annotations. 
In addition, framing constructs are assessed through individual article-level Likert prompts. 
Future work should compare these assessments with human annotations or domain-expert judgments, and should examine whether pairwise comparisons between political reference categories and triadic comparisons among left-, center- and right-oriented articles covering the same event produce different model-generated judgments.
Such comparative protocols would help examine whether judgments change when articles are evaluated in relation to one another rather than in isolation.


\paragraph{Ethical Considerations.}
As the study uses an LLM to assess framing constructs, we avoid treating model outputs as objective evaluations of news content. 
Instead, the analysis focuses on how these assessments vary across input conditions.
These precautions are especially important because the data collection concerns socially and politically consequential events, where over-interpreting model judgments could misrepresent coverage or reinforce simplified interpretations of complex public issues.

\begin{acks}
This research was funded in whole or in part by the Austrian Science Fund (FWF): \href{https://doi.org/10.55776/COE12}{10.55776/COE12},   \href{https://doi.org/10.55776/DFH23}{10.55776/DFH23}, \href{https://doi.org/10.55776/P36413}{10.55776/P36413}; 
and by the State of Upper Austria and the Federal Ministry of Education, Science, and Research: LIT-2024-13-SEE-111.
\end{acks}

\balance

\bibliographystyle{ACM-Reference-Format}
\bibliography{references}

@article{10.1111/j.1460-2466.2000.tb02843.x,
author = {Semetko, HA and Valkenburg, PM},
title = {Framing European politics: a content analysis of press and television news},
journal = {Journal of Communication},
volume = {50},
number = {2},
pages = {93-109},
doi = {https://doi.org/10.1111/j.1460-2466.2000.tb02843.x},
url = {https://onlinelibrary.wiley.com/doi/abs/10.1111/j.1460-2466.2000.tb02843.x},
year = {2000}
}

@inproceedings{arora-etal-2025-multi,
    title = "Multi-Modal Framing Analysis of News",
    author = "Arora, Arnav  and
      Yadav, Srishti  and
      Antoniak, Maria  and
      Belongie, Serge  and
      Augenstein, Isabelle",
    editor = "Christodoulopoulos, Christos  and
      Chakraborty, Tanmoy  and
      Rose, Carolyn  and
      Peng, Violet",
    booktitle = "Proceedings of the 2025 Conference on Empirical Methods in Natural Language Processing",
    month = nov,
    year = "2025",
    address = "Suzhou, China",
    publisher = "Association for Computational Linguistics",
    url = "https://aclanthology.org/2025.emnlp-main.1606/",
    doi = "10.18653/v1/2025.emnlp-main.1606",
    pages = "31531--31553",
    ISBN = "979-8-89176-332-6"
}

@inproceedings{tourni-etal-2021-detecting-frames,
    title = "Detecting Frames in News Headlines and Lead Images in {U}.{S}. Gun Violence Coverage",
    author = "Tourni, Isidora  and
      Guo, Lei  and
      Daryanto, Taufiq Husada  and
      Zhafransyah, Fabian  and
      Halim, Edward Edberg  and
      Jalal, Mona  and
      Chen, Boqi  and
      Lai, Sha  and
      Hu, Hengchang  and
      Betke, Margrit  and
      Ishwar, Prakash  and
      Wijaya, Derry Tanti",
    editor = "Moens, Marie-Francine  and
      Huang, Xuanjing  and
      Specia, Lucia  and
      Yih, Scott Wen-tau",
    booktitle = "Findings of the Association for Computational Linguistics: EMNLP 2021",
    month = nov,
    year = "2021",
    address = "Punta Cana, Dominican Republic",
    publisher = "Association for Computational Linguistics",
    url = "https://aclanthology.org/2021.findings-emnlp.339/",
    doi = "10.18653/v1/2021.findings-emnlp.339",
    pages = "4037--4050"
}

@inproceedings{wang-etal-2024-large-language-models-fair,
    title = "Large Language Models are not Fair Evaluators",
    author = "Wang, Peiyi  and
      Li, Lei  and
      Chen, Liang  and
      Cai, Zefan  and
      Zhu, Dawei  and
      Lin, Binghuai  and
      Cao, Yunbo  and
      Kong, Lingpeng  and
      Liu, Qi  and
      Liu, Tianyu  and
      Sui, Zhifang",
    editor = "Ku, Lun-Wei  and
      Martins, Andre  and
      Srikumar, Vivek",
    booktitle = "Proceedings of the 62nd Annual Meeting of the Association for Computational Linguistics (Volume 1: Long Papers)",
    month = aug,
    year = "2024",
    address = "Bangkok, Thailand",
    publisher = "Association for Computational Linguistics",
    url = "https://aclanthology.org/2024.acl-long.511/",
    doi = "10.18653/v1/2024.acl-long.511",
    pages = "9440--9450"
}

@inproceedings{chen-etal-2024-humans,
    title = "Humans or {LLM}s as the Judge? A Study on Judgement Bias",
    author = "Chen, Guiming Hardy  and
      Chen, Shunian  and
      Liu, Ziche  and
      Jiang, Feng  and
      Wang, Benyou",
    editor = "Al-Onaizan, Yaser  and
      Bansal, Mohit  and
      Chen, Yun-Nung",
    booktitle = "Proceedings of the 2024 Conference on Empirical Methods in Natural Language Processing",
    month = nov,
    year = "2024",
    address = "Miami, Florida, USA",
    publisher = "Association for Computational Linguistics",
    url = "https://aclanthology.org/2024.emnlp-main.474/",
    doi = "10.18653/v1/2024.emnlp-main.474",
    pages = "8301--8327"
}

@article{McInnes2018, doi = {10.21105/joss.00861}, url = {https://doi.org/10.21105/joss.00861}, year = {2018}, publisher = {The Open Journal}, volume = {3}, number = {29}, pages = {861}, author = {McInnes, Leland and Healy, John and Saul, Nathaniel and Großberger, Lukas}, title = {UMAP: Uniform Manifold Approximation and Projection}, journal = {Journal of Open Source Software} }

@article{VallejoVera2025,
  author    = {Sebastián Vallejo Vera and Hunter Driggers},
  title     = {LLMs as annotators: the effect of party cues on labelling decisions by large language models},
  journal   = {Humanities and Social Sciences Communications},
  year      = {2025},
  volume    = {12},
  number    = {1},
  pages     = {1530},
  doi       = {10.1057/s41599-025-05834-4},
  url       = {https://doi.org/10.1057/s41599-025-05834-4},
  issn      = {2662-9992},
  month     = sep,
  day       = {29}
}

\end{document}